# Multipeak Negative Differential Resistance from Interplay between Nonlinear Stark Effect and Double-Branch Current Flow

Mikołaj Sadek, Małgorzata Wierzbowska*, Michał F. Rode and Andrzej L. Sobolewski

*Institut of Physics, Polish Academy of Science (PAS),  Al. Lotników 32/46, 02-668 Warszawa, Poland*

*Multipeak negative differential resistance (NDR) molecular devices are designed from first principles. The effect of NDR is associated with the non-linear Stark shifts and the electron localization within the conductive region and contacts. Deep I(V)-curve well is formed when the aromatic molecule, containing intramolecular hydrogen bond, is connected to each lead by the double-branch contacts. This effect occurs at the same voltage where a single-junction case exhibits only a flat step in the current characteristics. The multipeak oscillations arise from the mutual effect of the Stark shifts located  at the electron-rich contacts and parts of the molecule – this opens the route for further tailoring the desired properties.*

___________________________________________________________________________

* Corresponding author, email: malwi45@gmail.com

# 1. Introduction

The negative differential resistance was discovered in 1958 by Japanese scientist Leo Esaki,[1] who was awarded the Nobel Prize in 1973 for tunneling phenomena. Applications of the NDR effect are very timing due to large progress in molecular electronics - they include resonant-tunneling diodes,[2] oscillators[3] operating even at terahertz frequencies,[4] amplifiers,[5] random-access memory devices and analog to digital signal converters.[6,7] Switching speed depends on the size of tunneling device, therefore molecular switches are especially desired. Well pronounced multipeak NDR could be used in the multiple-value memory and logic devices - however, finding a molecule with such characteristics is much more rare than a system with a single-dip in the I-V curve.

Single-valley NDR is nowadays a quite popular phenomenon, it has been found in various nanosystems, such as: atomic wires[8,9] and clusters,[10] molecular wires,[11-13] molecule-wire combinations,[9,14] molecular monolayers,[15-17] aromatic molecules and fullerenes,[18-22] DNA[23] and oligopeptides,[24] organic and metal-organic molecules,[11,12,15,17,25-31] magnetic molecules,[11,12,26,30,32,33] graphene field-effect transistors (FET),[34] Fe-doped graphene nanoribbons (GNR),[35] hybrid zigzag SiC-boronphosphide-SiC nanoribbon,[36] bulk GaAs[5] and ZnO nanoparticles.[37]

Depending on a system, a variety of mechanisms have been proposed to explain the drop of current under increasing voltage. First, the most intuitive descriptions of the NDR effect were associated with the junction breaking and contact or conductor or lead geometry.[8-10,12,24] Further, the chemical substitution,[20,23,27] pH of the conductor,[27] and polarization[18] were found to play a role. For the magnetic systems, either magnetization and the spin flip,[13,26,32,33] or the Coulomb blockade[32,38] and the spin blockade[32] are suspected to cause the non-monotonic I-V curve behavior. Finally, the orbital-energy position related phenomena were suggested: the conductor to the lead orbital coupling,[26] the linear Stark effect or the LUMO resonance shift with the bias,[14,28] a misalignment of the localized or the interface states,[29,39] an alignment between the lead states located around the Fermi energy with the lowest unoccupied molecular orbital of the central molecule.[22] For the molecules at the surfaces or the organic layers, the image-charge effect[17] or the trapping and releasing of the interface electrons under the light radiation[40] have been found to drive the NDR. For some highly symmetric systems, such as graphene FET structures, the band symmetry was responsible for the studied effect.[34] Similarly, the local orbital symmetry matching between the electrodes and the molecule was pointed.[15] Also the electron-hole binding across the molecule-electrode interface,[21] and the polaron formation[31] were examined as possible mechanisms of the NDR. Interesting mechanism based on the variation of the relative humidity was reported.[37,41] Hysteretic NDR effect was observed for the organic molecule connected by sulfur to metal leads and explained on the basis of slow charge capture - reduction or oxidation.[42] The strength of the NDR – the peak to valley ratio – can be attenuated by changing a composition of the junction.[43]

The multipeak-NDR circuits are usually built by a set of simpler electronic devices.[44,45,46] Only a few molecular-based nanosystems have been reported to possess this property.[30,40,47,48] However, their molecular structures are more complex than very simple molecules presented in this work.

We use the Wannier-based transport code,[49] with the input from the density functional theory,[50,51] and examine the conditions for the presence of multipeak NDR effect in a set of small aromatic molecules bound to the single-wall carbon nanotubes (SWCNT). The chemical group or atom used for the connection between the conducting molecule and the nanotubes, and the way of binding via a single or a double-branch arrangement, make great difference in the I-V characteristics. The molecules which exhibit the multipeak NDR effect - for their enol and keto tautomeric forms - have been demonstrated to possess the properties of good field transistors, when they were connected to Au leads via sulfur.[52] In this work, we associate the non-monotonic current behavior with the nonlinear Stark effect in the molecule and contacts. The multipeak NDR is a mutual effect of the charge polarization on the electron-rich contacts and parts of the molecule. The shape of the multivaley NDR-curve looks like a superposition of one deep well – of above 2:1 peak to valley ratio – and more shallow oscillating fine structure. Interestingly, presence of the neck-like part in the molecular structure also plays the role, since naphtalene and diphenyl molecules behave differently when connected to the leads in the same manner [vide infra]. For some choices of the contacts, the bias-voltage direction with respect to the enol-keto group decides whether the NDR effect occurs or not. Our theoretical studies may inspire further investigations of molecules and contacts to tailor multipeak NDR devices.

## 2. Results

We investigate transport in the molecules, 2-(pyridin-2-yl)phenol, composed of two phenyl rings and augmented with hydroxy/azine or oxo/amine moieties which are related to each other by proton transfer (PT) reaction along existing intramolecular hydrogen bond. The PT reaction occurs at certain critical values of the applied electric field, and switches the conductor between the two tautomeric forms.[52] The hydroxy/azine (enol) and the oxo/amino (keto) form of molecules considered in this work are presented as insets in Figure 1. At zero bias, the nominal (enol) structure is the stable one (with H connected to O). When the applied electric field increases (and its direction is parallel to the main molecular axis, with the negative side to molecular spot at the nitrogen atom) proton moves from the oxygen to the nitrogen atom, and keto structure is stabilized. Changing the direction of the external electric field, the PT process is reversed, although at different value of the field. Such hysteresis is different in the isolated molecule and in the molecule connected to leads, and depends also on the contacts and the way of the connection. We chose the atomic oxygen and $CH_2$ group for the contacts, and the carbon nanotubes CNT(6,0) or CNT(5,0) for the leads. When the molecule is isolated, the switching electric-fields – or the voltages corresponding to the electric field, and applied to the molecule ends are; 0.017 a.u. and 0.007 a.u. for enol-to-keto and keto-to-enol transitions, respectively. For the oxygen contacts as in Figure 1, the corresponding values of voltage are 0.015 a.u. and 0.006 a.u. The distance between the applied voltage ends is measured from the C atom at the source CNT to the C atom at the drain CNT). For the same geometry and $CH_2$ contacts, the field values are 0.013 a.u. and 0.004 a.u., respectively. When we connect the molecule directly to the CNT by a single-channel, the switching voltages are 0.017 a.u. and 0.006 a.u., respectively.

In all cases, the polarization of the electric-field is oriented with the negative sign at the nitrogen-side.

Our choice of the leads for the carbon nanotubes was dictated by good thermoelectric properties of the carbon-based systems, better than those of the metallic electrodes - which dissipate much of the energy for heat. The vibrational model and calculations for the local heating of the contacts under the applied bias were primarily reported for the metallic systems.[53] The vibrational contributions to the current, when the CNT leads were applied, also have been addressed.[54]

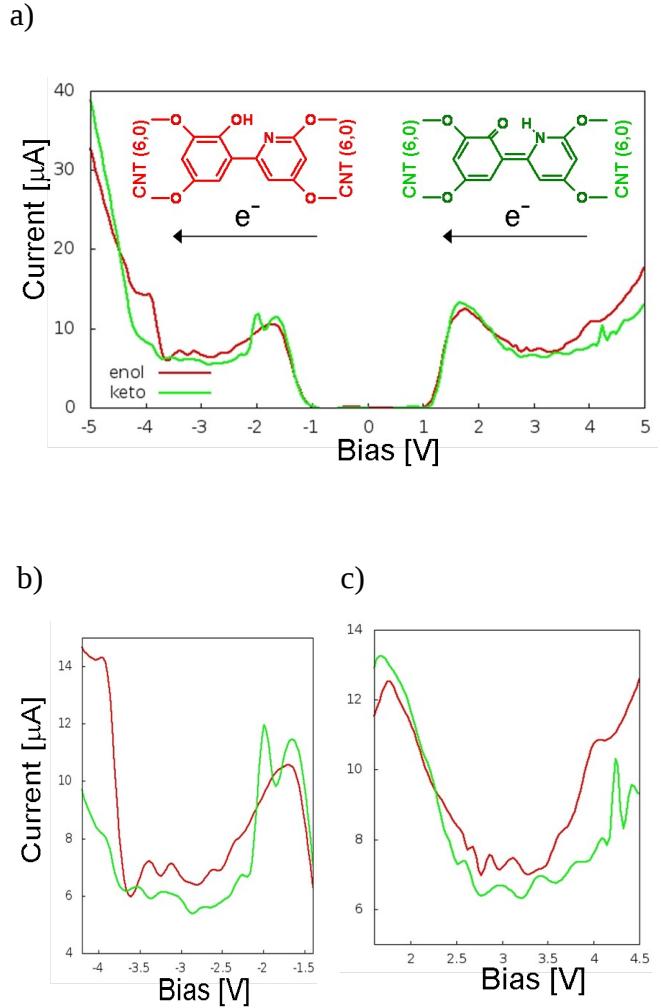

**Figure 1.** The I-V characteristics of tautomeric structures, i.e. enol (red) and keto (green) conductors, with double oxygen contacts to the CNT(6,0) leads.

In Figure 1 we present the current calculated for the enol and keto structures with the oxygen double-branch contacts to the CNT(6,0) leads. For both bias polarizations, we obtained wide (1-2V) and deep NDR valley with a "fine structure" of multiple peaks – the two lower panels with zoomed range show it more precisely. The peak-to-valley ratio of the wide valley for the positive voltage (2-3V) is about 2:1. The fine multipeak structure is damped with respect to the depth of the main valley, however, some twin features are still well pronounced. The shape of the giant valley and the multipeak structure depends on the bias polarization and the tautomeric forms (the enol and keto structures).

Although the similar molecules were proved to be very good transistors between the Au leads with the single-branch sulfur contacts,[52] in our case, when the NDR effect occurs, their I-V curves don't differ much in the current strength. It will be reported elsewhere that these molecules can be employed as transistors between the CNT leads when the single-branch connections are used in a conjunction with some choices of the non-metallic contacts.

Further in Figure 2, we present very similar NDR wells obtained for the $CH_2$ contacts applied to the enol and keto conductors. It is worth to note that the NDR does not occur in the enol case at the negative bias polarization. If the keto form would be stable, the region [-1V,-2V] could be used for transistor purposes, but this is not the case; for the negative bias, only the enol form is stable. Stabilization of this form for the negative bias can be obtained by proper chemical modification.

In order to check a role of hydrogen bond playing in the systems studied above, we decided to compare them to the diphenyl and naphthalene molecules connected to the CNT(6,0) via oxygen or $CH_2$ moiety in the same double-branch way; c.f. Figure 2. We did it only for the positive bias since the molecules are symmetric. These numerical experiments also showed the NDR effect. However for naphtalene, it depends on the contacts used - suggesting that removal of the neck-like structure of diphenyl also plays the role when it is combined with a respective type of the contacts.

Proceeding according the geometric suggestion, we checked the molecules with enol and keto structures connected directly to the CNT(5,0) via the single-branch bonds - see the I-V curves in the last panel in Figure 2. In this case the NDR did not show up. We checked that the same geometry with O-contacts and C-C-contacts also do not cause the NDR. Moreover, the direct-symmetric single-branch connection is less resistive than the double-branch connection via $CH_2$, therefore the current-switching properties of the device are worse, and the molecule orientation with respect to the bias polarization does not play a role.

Additionally, checking the literature, we found that the NDR effect was reported for many systems with double-branch, or even more, connections to the leads. Recently published, the combined double-single connections of Si clusters exhibit the NDR.[10]

In the same mixed way, the organic wires have been connected.[13] Due to the square molecular structures, Fe-phtalocyanines were connected with leads in the symmetric double-branch way.[26]

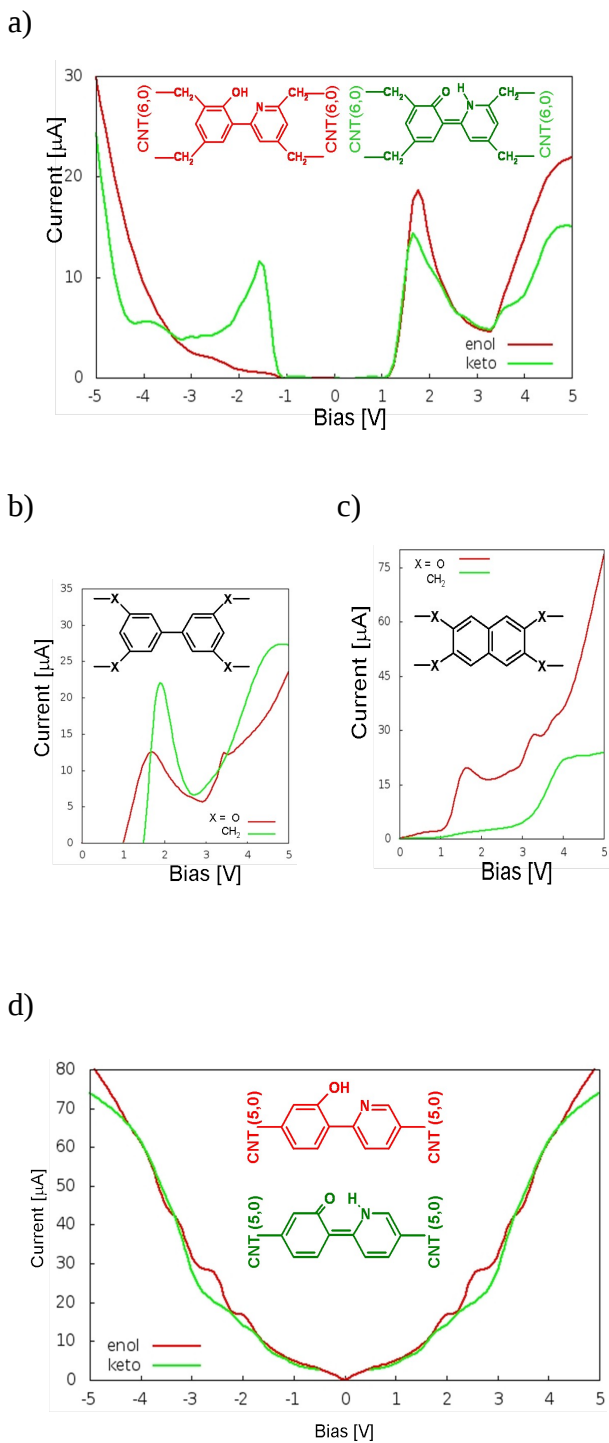

**Figure 2.** I-V characteristics of: (a) enol and keto conductors with $CH_2$ contacts to the CNT(6,0) leads, (b) diphenyl and (c) naphtalene with oxygen and $CH_2$ contracts to CNT(6,0) leads and (d) enol and keto molecules connected directly via the central single-bond to the CNT(5,0) leads.

Multiple-branch connections are present in nanorribons, and these systems also show the NDR.[34,35,36] The networks of organic molecules are multiple connected as well, and were reported as the NDR systems.

As it will be mentioned later in this section, the multiple connections probably strengthen the anisotropic response effects present in the triangular lattice; and therefore in the organic systems, as well.[40,48]

The electron-rich moieties contained in the conducting molecule, on the other hand, seem to be responsible for multiple current oscillations in our case; similar systems were reported as single-NDR[13] and multiple-NDR[48] devices as well.

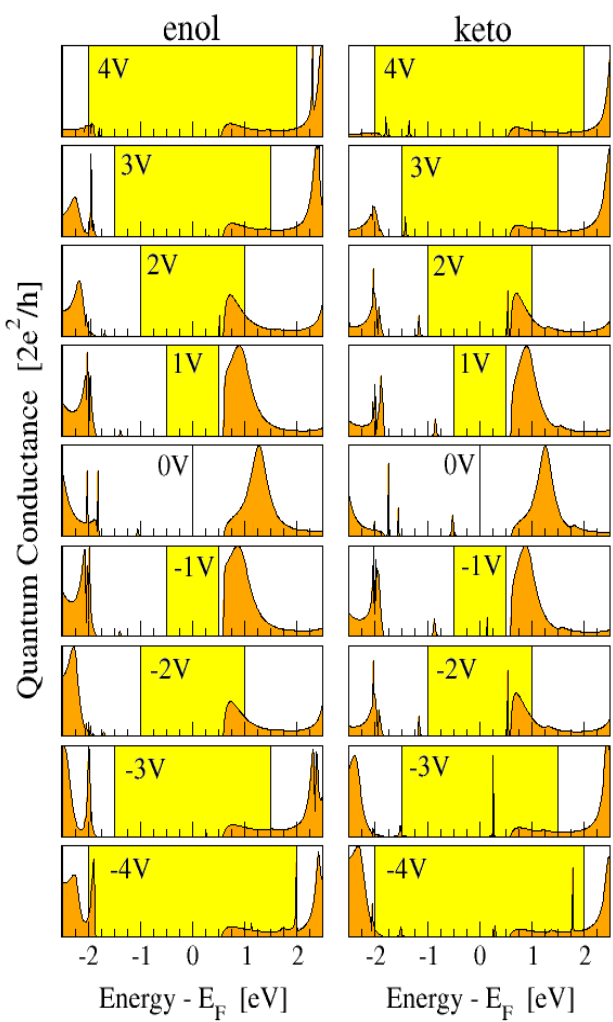

**Figure 3.** Quantum conductance in tautomeric forms, enol (left) and keto (right) structures, connected via oxygen double-channel to the CNT(6,0) leads; drawn for several voltages.

In order to clarify the reason for the negative differential resistance, we look closer to the calculated quantum conductance (QC) for the enol and keto structures connected via oxygen to the CNT(6,0) leads – c.f. Figure 3. We computed current at the applied voltage taking an integrand of the quantum conductance calculated "on-top" of the DFT Bloch functions within the bias range – as explained in the theoretical section. If the quantum-conductance function is constant with the applied voltage then the current obviously grows for higher voltage because of a wider range of the integration. But in our case, this function decreases with the external electric field, at the energies close to the Fermi level $E_F$. The superposition of the QC-decrease and the growing integrand-range causes the negative differential resistance when the first component is stronger. Around the bias of 3V, the QC function seems to be almost constant with the increasing applied voltage, and we again observe the growing current.

It is demanding to search for the reasons of such behavior of the QC function. Therefore, we plot the projected density of states (PDOS) as a function of the applied bias for the interesting parts of the investigated system – see Figure 4. The right side of the molecule is the one containing the hydroxy/oxo group.

Examining the PDOS pictures, it is clear that the Stark effect in our system is nonlinear. The most spectacular changes occur around +/-3.6V, but there are also the band anti-crossings around 2V, -2.3V, +/-2.5V. The substantial changes in the band degeneracies at voltages +/-3.6V coincide with the uprise of the current curve leaving the NDR region.

The PDOS pictures show very similar pattern for all fragments - which is, in fact, the fingerprint of the band structure of the whole system, and only the intensities depict the local structure. We keep in mind that when changing the bias polarization the atoms at source and drain are exchanged; i.e. when the bias is positive, the source is closer to nitrogen in the enol group, and for the negative bias, the source is closer to the O-enol side. Therefore swapping the bias polarization, the intensities of the source and the drain swap too. On the other hand, the intensities of the PDOS at the molecular O are similar for both current directions (equally far from the source and the drain).

Interestingly, existence of the non-linear Stark effect in our system should not be surprising, since we have a kind of triangular lattice (due to the sp2 hybridization at the carbon atoms), which was found to generate the anisotropic response to the electric field in many systems.[55]

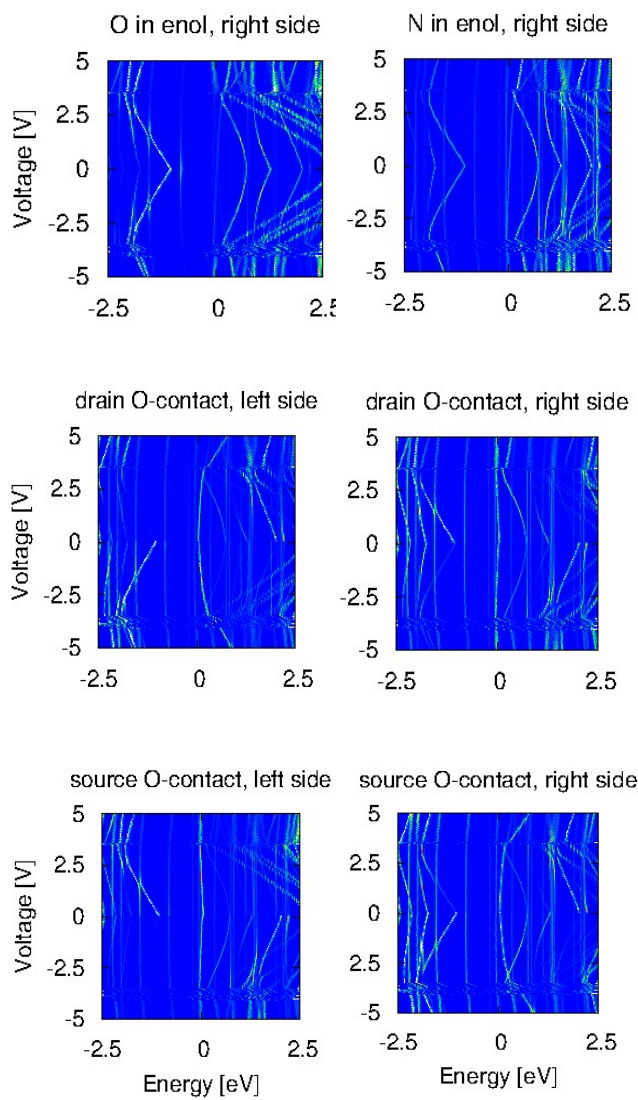

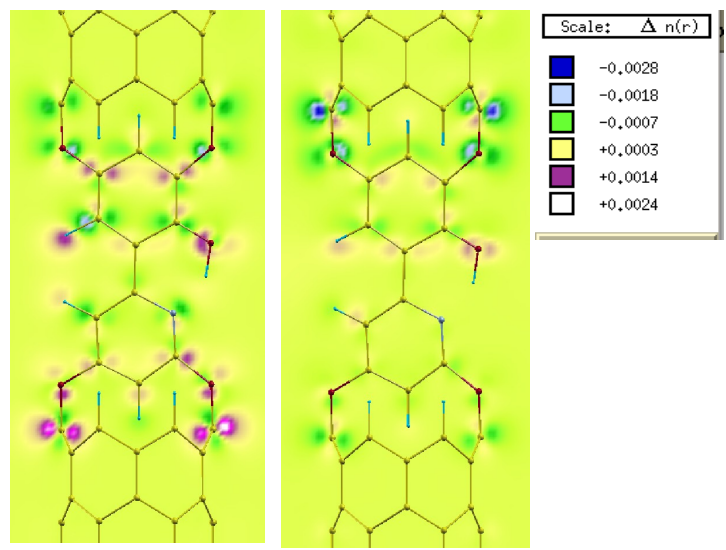

**Figure 5.** The difference of the electronic density Δn(r) for the system at two applied voltages ΔV=$V_2$-$V_1$ in two regions of the I-V curve: $V_1$=2V, $V_2$=2.5V (left) and $V_1$=3.75V, $V_2$=4.25V (right). The scale is identical for both cases.

**Figure 4.** The PDOS on the oxygen and nitrogen atoms in enol group and the oxygen atoms in the source and drain contacts to the CNT(6,0) leads; drawn for the applied voltages from -5V to 5V.

It is interesting to take a closer look at the electronic densities drawn under various external electric-fields; i.e. voltages between the CNT terminal carbons. In Figure 5, we present the differential densities in two bias regions: i) for the falling down current (2-2.5V) and ii) uprising current (3.75-4.25V). The positive values in this map indicate the electronic density; i.e the negative charge. The scale is the same for both cases and it is only about 0.5% of the total electronic density. Little in the amplitude, although very instructive, patterns tell as that in the current-falling case, the voltage growth causes the charge accumulation mainly in the source region, and less at the drain O-contacts and the enol's oxygen. At larger bias, leaving the NDR well, the charge deficit is visible in the drain region. The first effect blocks the electric flow through the molecule and the second sucks the current. The left- or right-side contacts, especially via oxygen, are very resistive because the coulomb blockade forms at the source. On the other hand, they better exhibit the current-switching effects originating from the tautomeric structures. The differential density pictures show also that for larger bias, after the charge accumulation at oxygens, the positive potential of the drain causes the charge deficit at the terminal carbons of the nanotube electrode. In the above effects, the relative electronegativity of the contacts and the electrodes plays a diverse role in the low and the high bias regimes.

## 3. Conclusion

In summary, we have predicted the existence of the multipeak negative differential resistance in the organic molecules, possessing the intramolecular hydrogen bond, connected to the carbon-nanotube electrodes. The shape of the I-V valley is wide and exhibits oscillating fine-structure with many smaller dips. The peak to valley ratio of the main well achieves as large value as 3:1. Detailed structure of the NDR depends on the proton switch and choice of the contacts. Electron rich atoms or groups are more promising for future modeling of the NDR. Although, the less efficient contacts may provide an another opportunity with respect to considered systems: to be useful as transistors when the NDR effect is modulated due to the proton-transfer process. Concluding, the NDR is caused by the nonlinear Stark effect in the molecule and contacts, and appears when the molecule is connected to the CNT via the double-branch bonds. Centrally and symmetrically single-branch connected enol- and keto-structures investigated in this work do not possess these properties. Our findings may open new perspectives for tailoring logic devices operating at terahertz regime.

## 4. Theoretical Section

Our calculations are based on the density functional theory[50] and Wannier functions.[56,57] First, we used the plane-wave Quantum ESPRESSO code (QE) for the self-consistent calculations of the molecules between the SWCNT. We did it for many discrete external electric fields with a step varying depending on the region of the I-V curve. When the further transport calculations showed a smooth curve, the step was equivalent to 0.25V applied to the terminal C-atoms of the SWCNT. In the most interesting regions this step was equivalent to 0.1V or even 0.05V.

In the second step, we used the wannier90-2.0.0 code[58] to obtain the maximally-localized Wannier functions for the whole system, and then to perform the transport calculations within the Landauer-Bűttiker scheme.[59] Obtained this way quantum conductance (or transmission) for each electric field separately, $T(\varepsilon; E \sim V)$, was embedded in the equation[60]

$$I(V) = \int [f(\varepsilon - \varepsilon_F + V/2) - f(\varepsilon - \varepsilon_F - V/2)] \, T(\varepsilon; E \sim V) \, d\varepsilon,$$

where the Fermi-Dirac distribution $f(\varepsilon)$ was used.

Then, we found the current (I) at the bias (V) equivalent to the external electric field (E) used in the corresponding DFT calculations. At the end, the calculated points were interpolated to obtain the full I-V curve.

In order to estimate the voltage, at which the transistor switching between the enol and keto structures occurs, we minimized the total energy of a given tautomeric structure by optimizing geometry of nuclei at discrete values of the applied electric field (with the step of 0.001 a.u.), using the DFT-based code of TURBOMOLE[61] with the B3LYP functional and the correlation-consistent valence double-zeta gaussian basis set with polarization function for all atoms (cc-pVDZ).[62] This procedure was done for each contact presented in this work, and SWCNT were simulated with the terminating $CH_3$ group.

To complete the technical information, we give details of the input setup: for the DFT-transport calculations the BLYP functional was used, the energy cutoff for the plane-waves in the Quantum Espresso code was set to 30 Ry, and the pseudopotentials were of the Martin-Troulliers type. In the QC calculations, each lead – left and right – was built by two CNT units, and the conductor region consisted of the molecule with the contacts and one CNT ring saturated with hydrogens on both sides. We used the CNT(6,0) and CNT(5,0) for the simulations of double- and single-branch connections, respectively.

The PDOS pictures were prepared with the electric-field discrete step corresponding to the voltage of 0.1V. Figure 5 have been prepared with the XCrySDen package.[63]

*Acknowledgments*

*We thank Michael Thoss for the discussions in the course of the study. This work has been supported by the research projects of the National Science Centre of Poland, Grants No. 2011/01/M/ST2/00561 and 2012/04/A/ST2/00100. Calculations were done in the Interdisciplinary Centre of Mathematical and Computer Modeling (ICM) of the University of Warsaw within the computation grants G47-7 and G56-32.*